\documentclass{aa}  
\usepackage{graphicx}
\usepackage{times}
\usepackage{epsfig}
\usepackage{longtable}
\usepackage{lscape}
\usepackage{amssymb}
\usepackage{latexsym}
\usepackage{natbib}
\usepackage{txfonts}
\def\kms{$\rm km\;s^{-1}$}

\def\kmsp{$\rm km\;s^{-1}\;pixel^{-1}$}

\def\hb{H$\beta$}

\def\mgill{Mg~{\small I}($\lambda\lambda5167,5173,5184$)}
\def\mgi{Mg~{\small I}}

\def\fee{Fe~{\small I}($\lambda\lambda 5270, 5328$)}

\def\mguno{Mg$_1$}
\def\mgdue{Mg$_2$}
\def\mgb{Mg~{\it b}}
\def\fei{Fe{\small 5270}}
\def\feii{Fe{\small 5335}}
\def\feiii{Fe{\small 5105}}
\def\feiv{Fe{\small 5406}}

\begin{document}
\title{Kinematics and line strength indices in the halos of the Coma
  Brightest Cluster Galaxies NGC 4874 and NGC 4889 \thanks{Based on
  data collected at Subaru Telescope, which is operated by the
  National Astronomical Observatory of Japan.  }}

   \subtitle{}

   \author{L. Coccato \inst{1}
	  \and M. Arnaboldi \inst{2,3}
          \and O. Gerhard \inst{1}
	  \and K. C. Freeman \inst{4}
          \and G. Ventimiglia \inst{1,2}
          \and N. Yasuda \inst{5}
          }

   \offprints{L. Coccato, e-mail: lcoccato@mpe.mpg.de}

   \institute{Max-Plank-Institut f\"ur Extraterrestrische Physik,
     Giessenbachstra$\beta$e 1, D-85741 Garching bei M\"unchen, Germany.
     \and European Southern Observatory, Karl-Schwarzschild-Stra$\beta$e 2,
          85748 Garching bei M\"unchen, Germany.
     \and INAF, Osservatorio Astronomico di Pino Torinese, I-10025 Pino Torinese, Italy.
     \and Research School of  Astronomy \& Astrophysics, ANU, Canberra, Australia.
     \and Institute for the Physics and Mathematics of the Universe,
          University of Tokyo, 5-1-5 Kashiwa-no-ha, Kashiwa, Chiba
          277-8568, Japan.
}

   \date{\today}

% \abstract{}{}{}{}{} 
% 5 {} token are mandatory
 
  \abstract
   {}
   {We investigate the stellar kinematics and line strength indices in
     the outer halos of brightest cluster galaxies (BCGs) in the Coma
     cluster to obtain the outer halo $V_{\rm rot}$ and $\sigma$
     profiles and to derive constraints on the formation history of
     these objects.}
   {We analyzed absorption lines in deep, medium-resolution, long-slit
     spectra in the wavelength range $\sim 4500 - 5900$ \AA, out to
     $\sim 50$ kpc for NGC 4874 and $\sim$ 65 kpc for NGC 4889,
     probing regions with a surface brightness down to $\mu_R \sim 24$
     mag arcsec$^{-2}$.}
   {These data provide stellar velocity and velocity dispersion
     profiles along the major axes of both BCGs, and also along the
     minor axis of NGC 4889. The kinematic properties of NGC 4874 and
     NGC 4889 halos extend the previous relations of early-type galaxy
     halos to bright luminosities and indicate that the stars in the
     outer regions are still bound to these galaxies. For NGC 4889 we
     also determine \hb\, Mg and Fe line strength indices, finding
     strong radial gradients for Mg and Fe. The current dataset for
     NGC 4889 is one of the most extended in radius, including both
     stellar kinematics $and$ line strength index measurements. }
   {}

   \keywords{galaxies: cluster: general --
     galaxies: cluster: individual: Coma cluster -- galaxies:
     individual: NGC 4874 and NGC 4889 -- galaxies: kinematics and dynamics -- Galaxies: abundances }

   \titlerunning{Kinematics and line strength indices in BCG halos.}

   \authorrunning{Coccato et al.}

   \maketitle
%
%________________________________________________________________

\section{Introduction}

Brightest cluster galaxies (BCGs) are large and luminous galaxies
located in the centers of galaxy clusters. The formation history of
BCGs and their halos is connected to the formation of the cluster
itself (\citealt{Dubinski98}) and to the presence of diffuse
intra-cluster light (\citealt{Gonzalez+05, Murante+07}).
The formation of BCG and their halos can be investigated by the
combined study of the stellar kinematics and the population content
of their halos. Dynamical timescales in the halos are on the order of
1 Gyr \footnote{Estimated for distance $R=50$ kpc and circular
  velocity $V_C=250$ \kms, e.g. \citealt{Binney+87}.}, approaching a
significant fraction of the age of the universe, and therefore the
fingerprints of the formation processes may still be preserved there.

So far, the studies that provide both stellar kinematics and line
strength indices in BCGs have been limited to within one effective radius
(e.g \citealt{Carter+99}; \citealt{Fisher+95}; \citealt{Brough+07};
\citealt{Spolaor+08a, Spolaor+08b}; \citealt{Loubser+08, Loubser+09}),
therefore measurements over a wider radial range are highly
desirable.

This work is the first of a series aimed at studying the formation
history of BCGs by probing the stellar kinematics and populations of
their outer halos, covering regions at 3 effective radii or
larger. As first targets, we selected NGC 4874 and NGC 4889, the two
BCGs in the Coma cluster (Abell 1656).
The inner parts of these galaxies have been studied with photometric,
kinematic and stellar populations data documented and available in the
literature (e.g. \citealt{Mehlert+00, Gavazzi+03, Gerhard+07,
  Trager+08}).

In this paper we describe the data acquisition, reduction, and the
measurements of the long slit stellar kinematics for both  galaxies,
and the line strength indices for NGC 4889. The data set in NGC 4889
extends out to 65 kpc (which correspond to $\sim 4.3$ effective radii,
\citealt{Jorgensen+95}), providing  the most radially extended
measurements of absorption line kinematics {\it and} line strength
indices in the outer halo of a BCG.  These data are the basis of
forthcoming papers investigating the formation history and
evolution of the galaxies in the Coma cluster core.

The paper is organized as follows. Spectroscopic observations and data
reduction are discussed in Section \ref{sec:observations}. The sky
subtraction and the radial binning of the spectra are described in
Sections \ref{sec:sky} and \ref{sec:radial_binning},
respectively. Section \ref{sec:stellar_kinematics} describes the
measurements of the stellar kinematics, while Section
\ref{sec:indices} describes the measurements of the line strength
indices. Finally, the results are discussed in Section
\ref{sec:discussion}.

In this paper, we adopt a distance to NGC 4784 of $D=102.6$ Mpc and an
effective radius of $R_e = 70\farcs79 = 35.21$ kpc; for NGC 4889 we
adopt $D=92.7$ Mpc and $R_e = 33\farcs88 = 15.23$ kpc. Distances are taken
from the NASA/IPAC Extragalactic Database (NED) and the effective
radii from \citet{Jorgensen+95}.

\section{Spectroscopic observations and data reduction}
\label{sec:observations}

\begin{figure}
  \hspace{0.9cm}
\vbox{
 \psfig{file=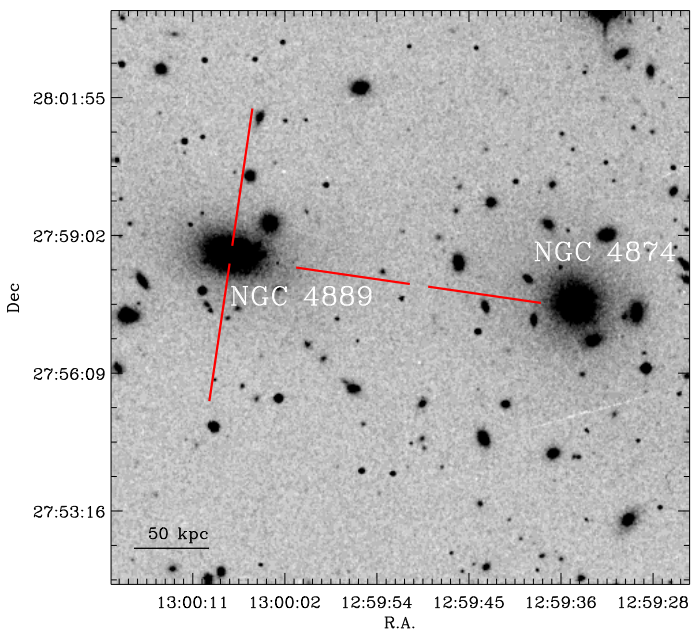,width=7.6cm}
% /home/lcoccato/COMA/data_paper/figures_long_slit/mkfield.pro}
}
\bigskip
\caption{Optical DSS image of the Coma cluster core. North is up, East is
  Left. The two red  lines with the central gap represent the portion 
  of the slit covered by the 2 CCDs in the FOCAS spectrograph. 
  The scale in kpc given on the bottom left
  corner is computed assuming a distance of 97.7 Mpc, which is the
  average of the distances to NGC 4889 and NGC 4874 as given by NED.
}
\label{fig:field}  
\end{figure}

Long slit spectra were acquired with the Faint Object Camera And
Spectrograph (FOCAS, \citealt{Kashikawa+02}) at the SUBARU telescope
of the National Astronomical Observatory of Japan (NAOJ), on Mauna Kea
(Hawaii, USA). Data were collected during two runs; in run \#1 (April
2007) we obtained 8 hours integration long slit data on the West side
of the NGC 4889 major axis, in the region in between NGC 4874 and NGC
4889, with a spectral resolution of 76 \kms.  In run \#2 (May 2008) we
obtained 5.5 hours integration of long slit data for the NGC 4889 minor
axis, with a spectral resolution of 96 \kms.  In Figure \ref{fig:field}
the observed field and slits locations are shown. In
Table \ref{tab:setup} we provide a summary of the observing log and
instrumental set up.
During these runs, we observed a set of kinematic template stars and
Lick spectrophotometric standard stars for calibration to the Lick
system.  Long slit spectra on blank fields of the COMA cluster were
also obtained for the sky background evaluation and to correct for
large-scale illumination patterns due to slit vignetting. In addition,
we observed at least one flux standard star per night, to flux
calibrate the spectra.

\begin{table}
\centering
\caption{Summary of the observing log and instrumental set up.}
\begin{tabular}{l c c }
\noalign{\smallskip}
\hline
Parameter           &   Run 1                 &    Run 2           \\
                    &   13-14 Apr. 2007       &    8-9 May. 2008   \\
\noalign{\smallskip}
\hline
\noalign{\smallskip}
Exp. time on galaxy &   $16\times 30$ min     &  $11\times 30$ min  \\
Exp. time on sky    &   $6\times 30$ min      &  $4\times 10$ min   \\
Mean Seeing         &      $\sim 2"$          &     $\sim 2"$       \\
R.A. slit center    &   12:59:51              &  13:00:08           \\
DEC. slit center    &  +27:58:01              & +27:58:37           \\
Slit P.A.           &  $81^\circ7$            & $-8^\circ3$          \\
Slit length         &  300 arcsec             & 300 arcsec          \\
Slit width          &  0.8  arcsec            &  1.0 arcsec         \\
CCDs                & MIT  2.2K $\times$ 4K   & MIT  2.2K $\times$ 4K\\ 
Gain                & 2.1 e$^-$ / ADU &2.1 e$^-$ / ADU \\
R.O.N.              & 4 e$^-$ rms             &4 e$^-$ rms           \\
Pixel size          & 15 $\mu$m               &15 $\mu$m             \\
Pixel binning       & $3\times1$              &$3\times1$            \\
Pixel scale         & 0\farcs104 pixel$^{-1}$ ($\times 3$) &0\farcs104 pixel$^{-1}$ ($\times 3$) \\
Grism               & VPH 450                 & VPH 520               \\
Filter              & L600                    & L600                 \\
Dispersion          &  0.394 \AA\ pixel$^{-1}$ ($\times 1$)&0.400 \AA\ pixel$^{-1}$ ($\times 1$)\\
Observed range      & 4200 -- 5700 \AA        &4450 -- 6050  \AA     \\
Calibrated range    & 4545 -- 5665 \AA        &4700 -- 5900  \AA     \\
Calibration lamp    & Th + Ar                 &Th + Ar               \\
Instrumental $\sigma^{(*)}$& 76 \kms\  & 96  \kms\ \\
\noalign{\smallskip}
\hline
\label{tab:setup}
\end{tabular}
\begin{minipage}{9 cm}
Notes -- $^{(*)}$ Computed at 5100 \AA.
\end{minipage}
\end{table}

Standard data reduction (bias subtraction, flat fielding to correct
for pixel to pixel chip sensitivity fluctuations) were performed
using standard {\small IRAF} \footnote{{\small IRAF} is distributed by
  NOAO, which is operated by AURA Inc., under contract with the
  National Science Foundation. } routines. Spectra distortion due to
the optics were corrected using {\small
  FOCASRED}\footnote{Information about the {\small FOCASRED} package
  can be obtained at the web page:
  http://www.naoj.org/Observing/Instruments/FOCAS/Detail/
  UsersGuide/DataReduction/focasred.html} data reduction package
within the {\small IRAF} environment.
Cosmic rays were detected and removed using {\small LACOS}
\citep{vanDokkum01}. Residual cosmic rays were removed by manually
editing the spectrum.
Wavelength calibration was performed using standard IRAF
routines. Comparison Th+Ar spectra were obtained during the night
close in time to the scientific spectra.  Because two different grisms
were used for the two runs, we optimized the wavelength calibration
in different spectral ranges: $4545 \AA < \lambda < 5665 \AA$, for run 1, and
$4700 \AA < \lambda < 5900 \AA$, for run 2. With this choice, we were
able to minimize the errors in the wavelength calibration, which are
about $\pm 10$ \kms\ as measured from the calibrated comparison
spectra and from the sky line emission lines. Absorption lines as \hb,
\mgill\ triplet and \fee\ fall inside the two selected wavelength
ranges.
Finally, the spectra were corrected for change in the slit
illumination along the slit direction using the spectra obtained on sky
blank fields.

\section{Sky subtraction}
\label{sec:sky}

Sky subtraction may introduce errors in the measurements from spectra
of low surface brightness regions. To limit systematic effects as much
as possible, we account for the sky subtraction in two ways.

The first method is to obtain sky spectra from blank fields. The sky
spectrum is then free of contamination either from the galaxy halos
or intracluster light, but the spectra are not taken simultaneously
with the scientific exposures. Therefore, if the relative intensities
of the sky continuum, emission or absorption lines change with time,
 sky residuals are present in the galaxy spectra.

\begin{figure}
%/home/lcoccato/COMA/data_paper/figures_long_slit/sky_check/compare_sky2.pro
\vbox{
 \psfig{file=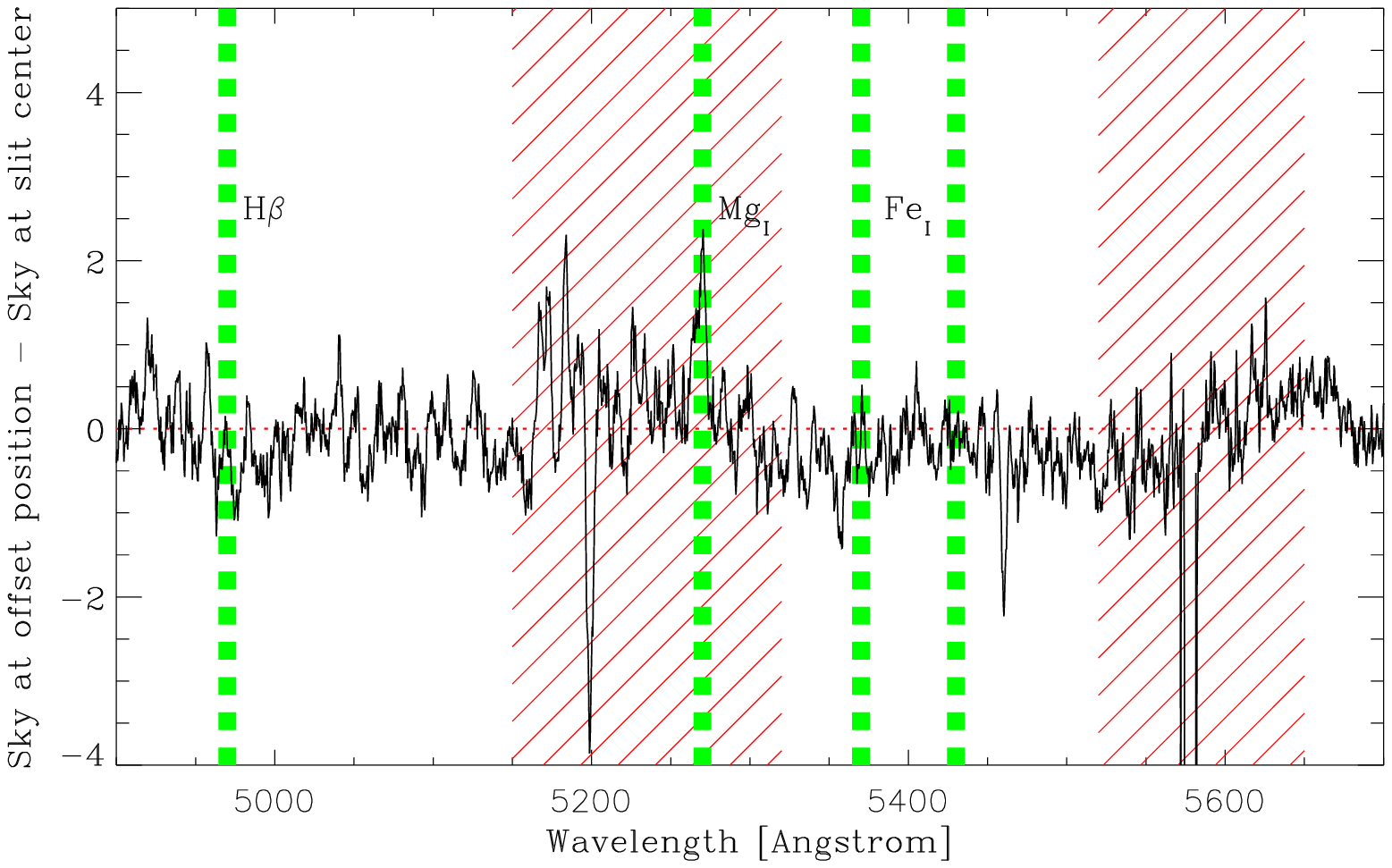,width=9.0cm,clip=}
 \psfig{file=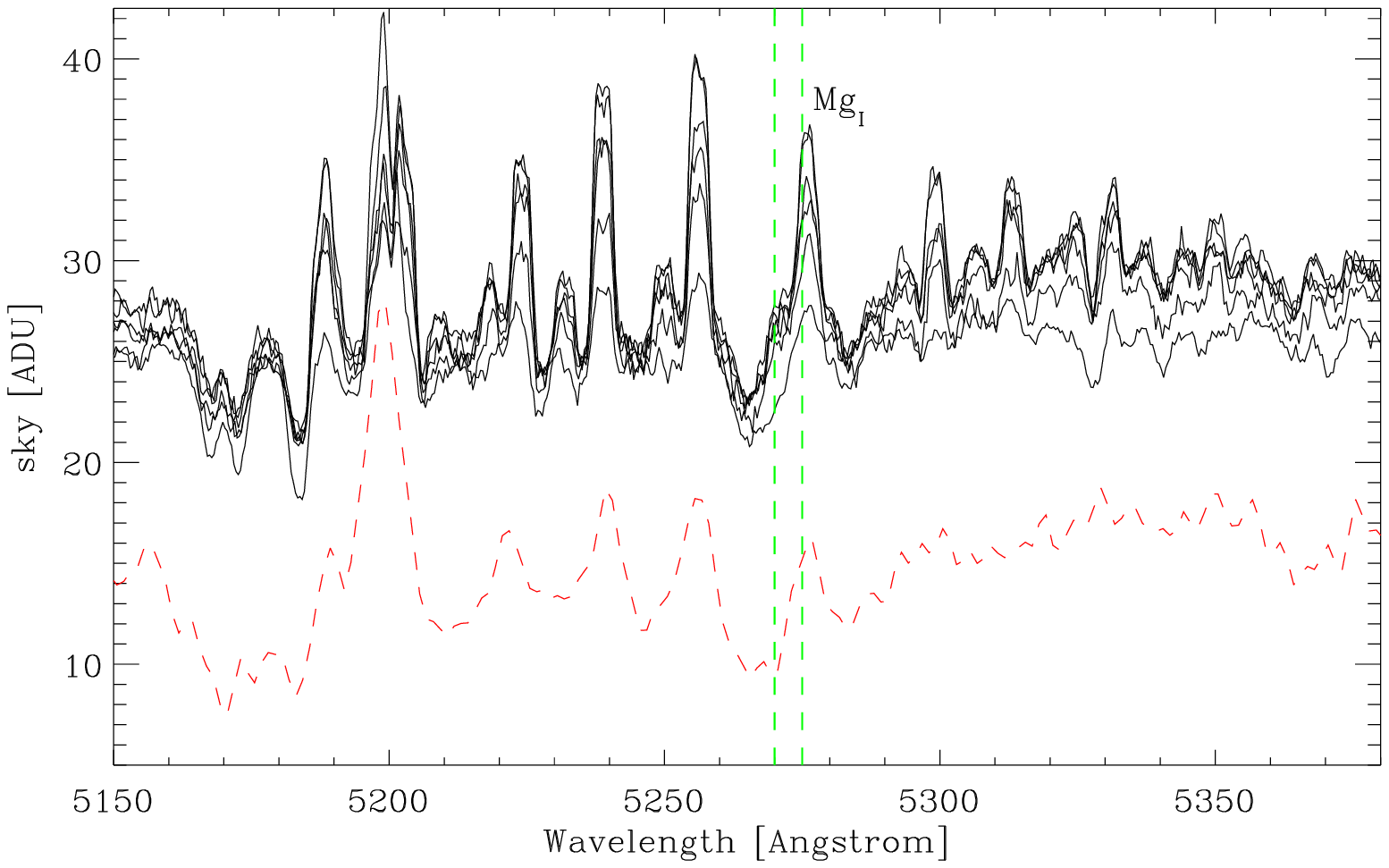,width=9.0cm,clip=}
}
\caption{{\it Upper panel}: difference between the mean sky spectrum
  measured at the offset position on the blank field and the sky
  spectrum measured at the long-slit center (run 1). The mean value of
  the difference is $\leq 1$ ADU. The positions where galaxy
  absorption line features should be visible (taking into account the
  galaxy radial velocity) are marked in green (see text for
  details). The spectral ranges containing the most intense sky
  emission lines are shaded in red. {\it Lower panel}: comparison of
  sky spectra obtained at different times during the night ({\it black
    continuous lines}). The time variation of the relative intensity
  of sky emission and absorption lines is clearly visible. As in the
  upper panel, the position of the galaxy \mgi\ absorption line is
  marked in green. {\bf For} comparison, we show also the typical sky
  spectrum visible from the Kitt Peak National Observatory ({\it red
    dashed line}, rescaled to allow comparison, from
  \citealt{Massey+00}). The same OH emission lines are visible.}
\label{fig:sky_comparison}
\end{figure}

The second method is to extract the sky spectrum from a region of the
slit where the galaxy light is negligible (central part of the slit
for run 1, outer borders of the slit for run 2). The sky is now
observed simultaneously to the galaxy observations but the
disadvantage is that it might contain a residual contribution from the
galaxy halo and/or intracluster light. We then may subtract spectral
features, which belong to the galaxy and which we are, in fact,
interested in measuring.

In the upper panel of Fig. \ref{fig:sky_comparison} we compare the sky
spectra extracted using the two different methods. If the sky spectrum
extracted at the slit center (run 1) or at the slit edges (run 2)
contains a small residual contribution from the galaxy (stellar
continuum and/or absorption line features) it should become visible
when comparing with the sky spectrum extracted at the offset
position. Residuals are approximately $\pm 0.5$ counts
(i.e. $\sim 2$\% of the average sky level at 5100 \AA). This means
that the level of the sky continuum computed with the two methods is
the same, and differences between the two methods (if any) should be
associated with spectral line features. However, we clearly see
that larger residuals are present in the wavelength regions of the
most intense sky lines. In particular, a residual spectral feature is
seen around $\sim 5280$ \AA\, which is very close to the position of
the \mgill\ triplet for the systemic velocity of NGC 4889. In the
lower panel of Fig.  \ref{fig:sky_comparison} we compare the sky
spectra extracted at offset position at different times during the
night. We concentrate on the spectral region from 5150\AA\ to 5380\AA,
which contains some OH emission lines \citep{Osterbrock+96}. The
variation of sky lines intensities over time is clearly visible and
its amplitude ($\sim 1-2$ AUDs) is consistent with the intensity of
the spectral feature observed at $\sim 5280$ \AA .  These time
variations indicate that the observed residual at $\sim 5280$ \AA\ is
related to the sky and can be best corrected for by using the sky
observed simultaneously to the scientific observations.

\section{Extraction of spectra in low surface brightness regions}
\label{sec:radial_binning}

Before measuring the stellar kinematics (Section
\ref{sec:stellar_kinematics}) and the line-strength indices (Section
\ref{sec:indices}), we needed to bin spatially the spectra along the
slit direction to increment the signal to noise ratio. Spectra from
adjacent columns\footnote{The spectra were oriented to have the
  dispersion direction along the vertical axis, and the slit aligned
  with the horizontal axis.}  were added in order to reach a minimum
signal to noise $S/N \sim 10$ for run 1, and $S/N \sim 20$ for run 2,
which ensure reliable measurements of stellar kinematics and line
strength indices.  For run 1, the extracted spectra have an overall
lower $S/N$ than run 2 because the slit was sampling regions of lower
surface brightness, and was illuminating regions of the CCD with bad
cosmetics and hot columns which required masking.

\begin{figure}
\psfig{file=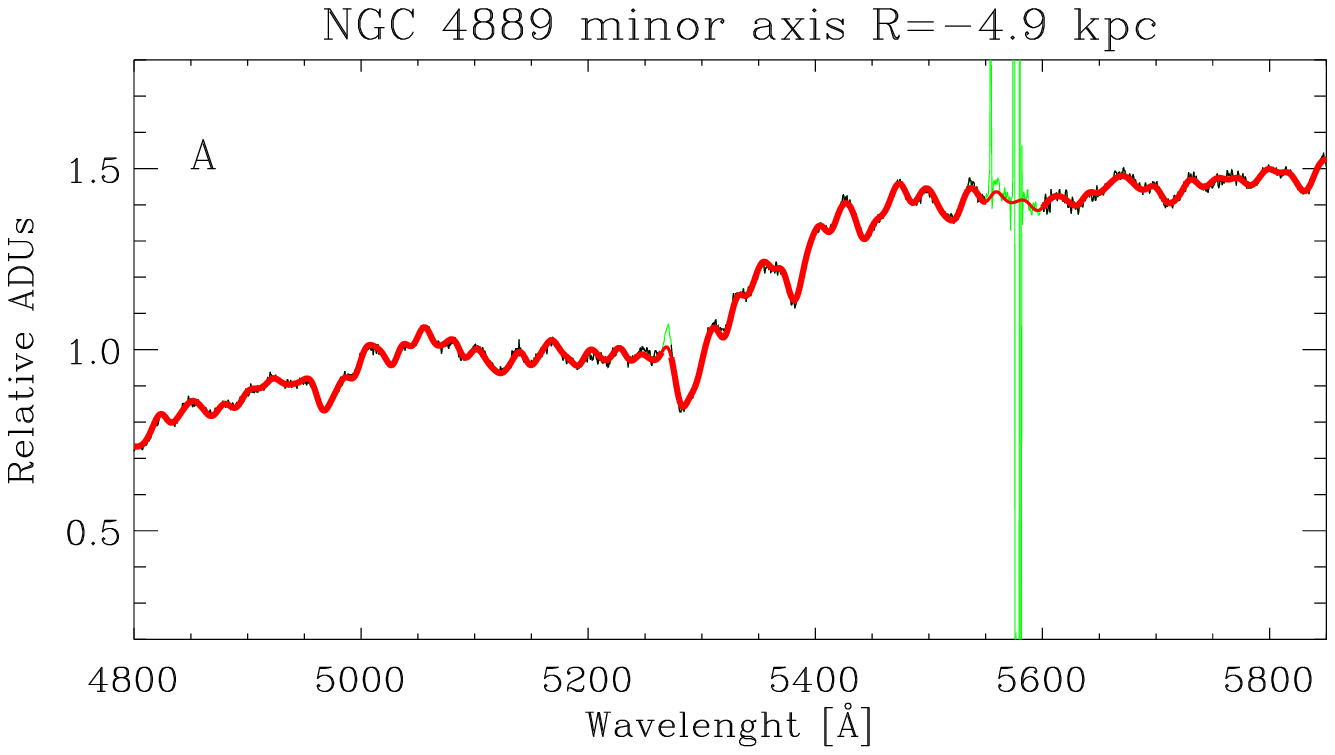,width=9.0cm,clip=}
\psfig{file=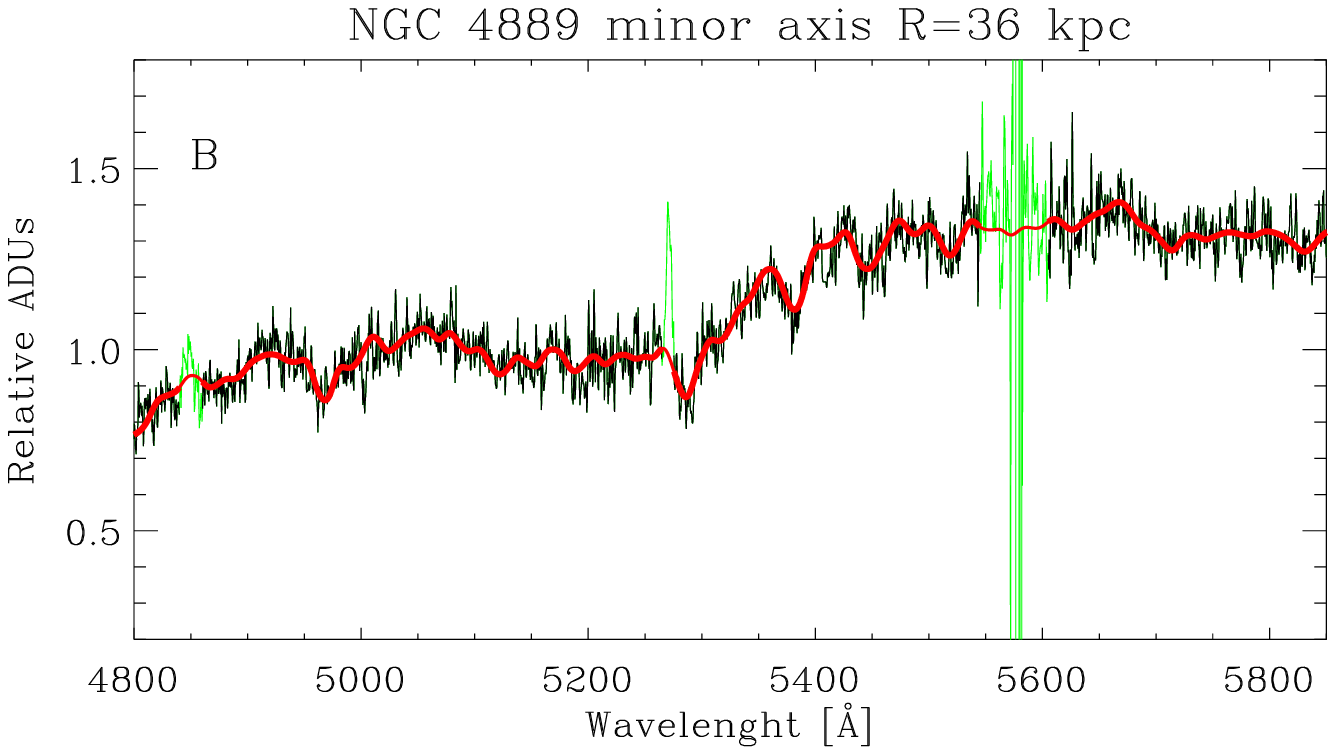,width=9.0cm,clip=}
\psfig{file=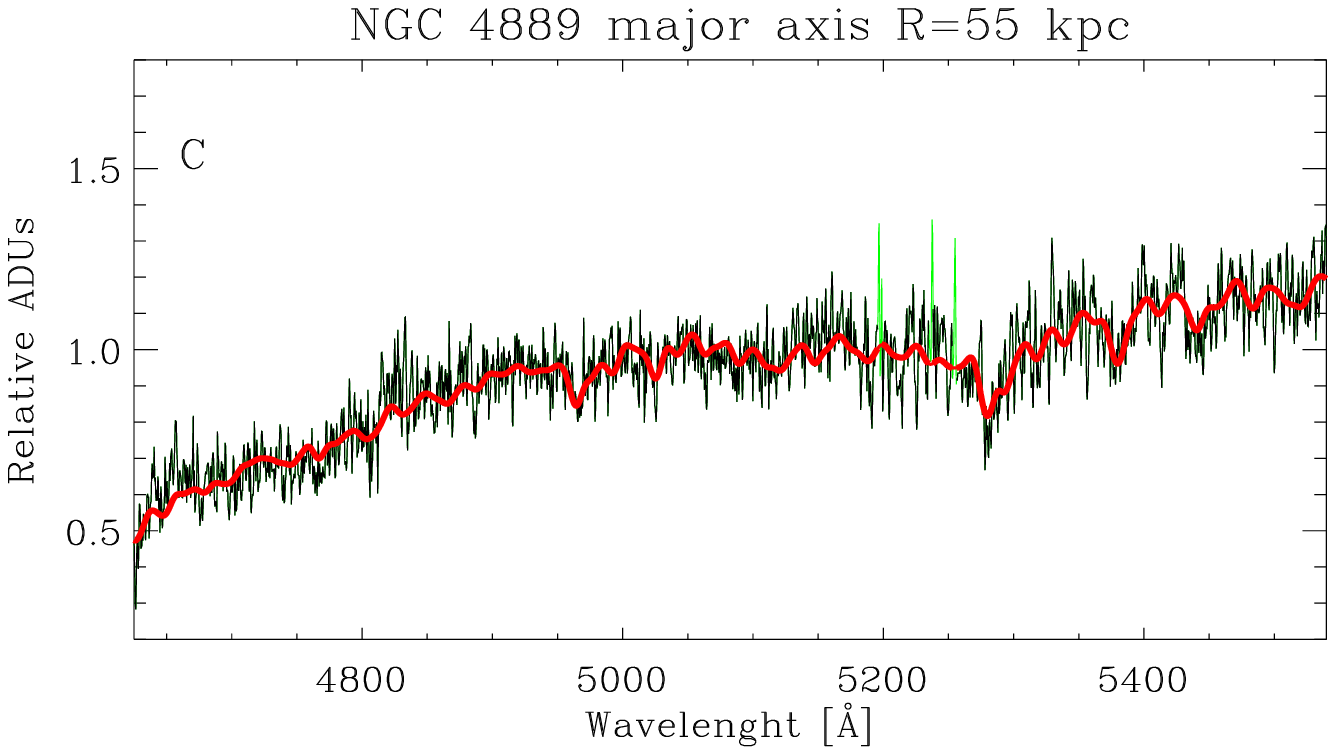,width=9.0cm,clip=}
\psfig{file=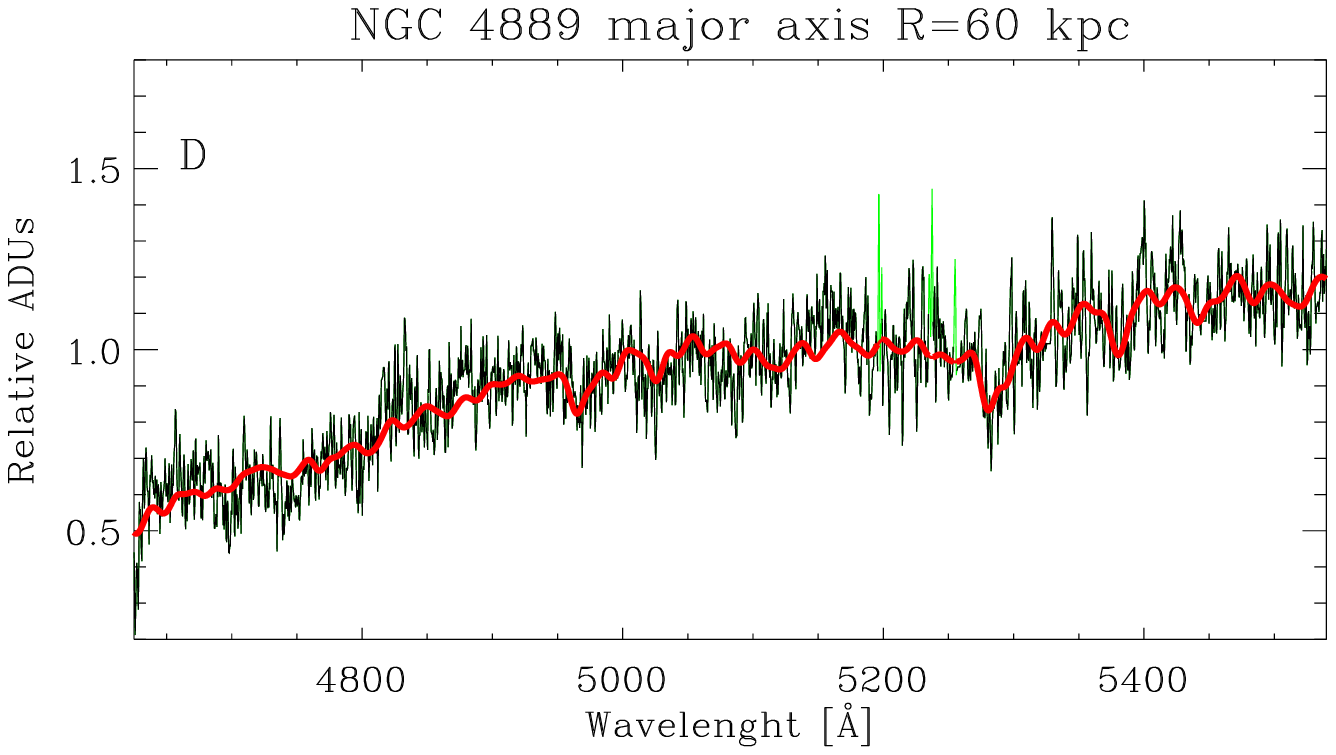,width=9.0cm,clip=}
\caption{Examples of kinematic fits. {\it Black:} galaxy
    spectrum, {\it green:} portion of galaxy spectrum excluded in the
    fit, {\it red:} best fit template model. The S/N ratio per pixel of the
    spectra in panels from A to G are: 61, 15, 13, 10, 9, 8, 8,
    respectively. All spectra are normalized to the value of the best
    fit model at 5100 \AA.}
\label{fig:fit1} 
\end{figure}
\addtocounter{figure}{-1}
\begin{figure}
\psfig{file=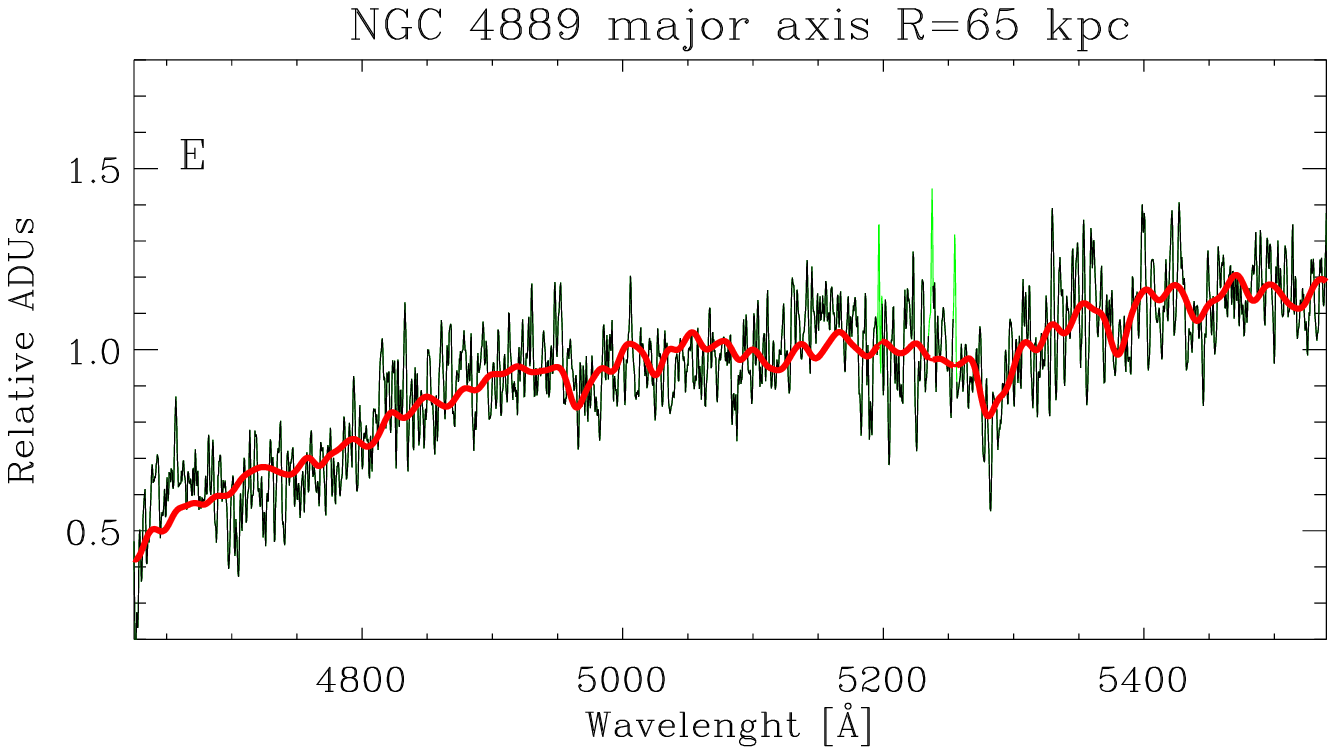,width=9.0cm,clip=}
\psfig{file=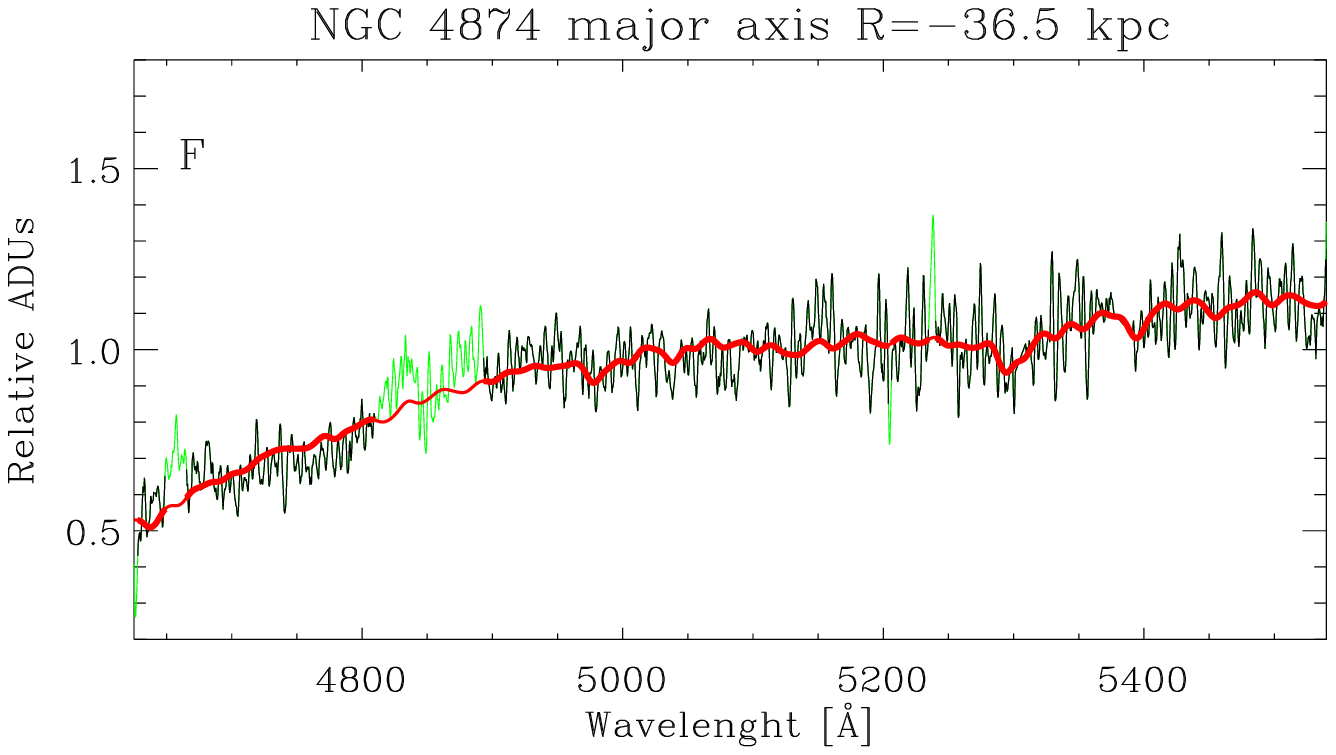,width=9.0cm,clip=}
\psfig{file=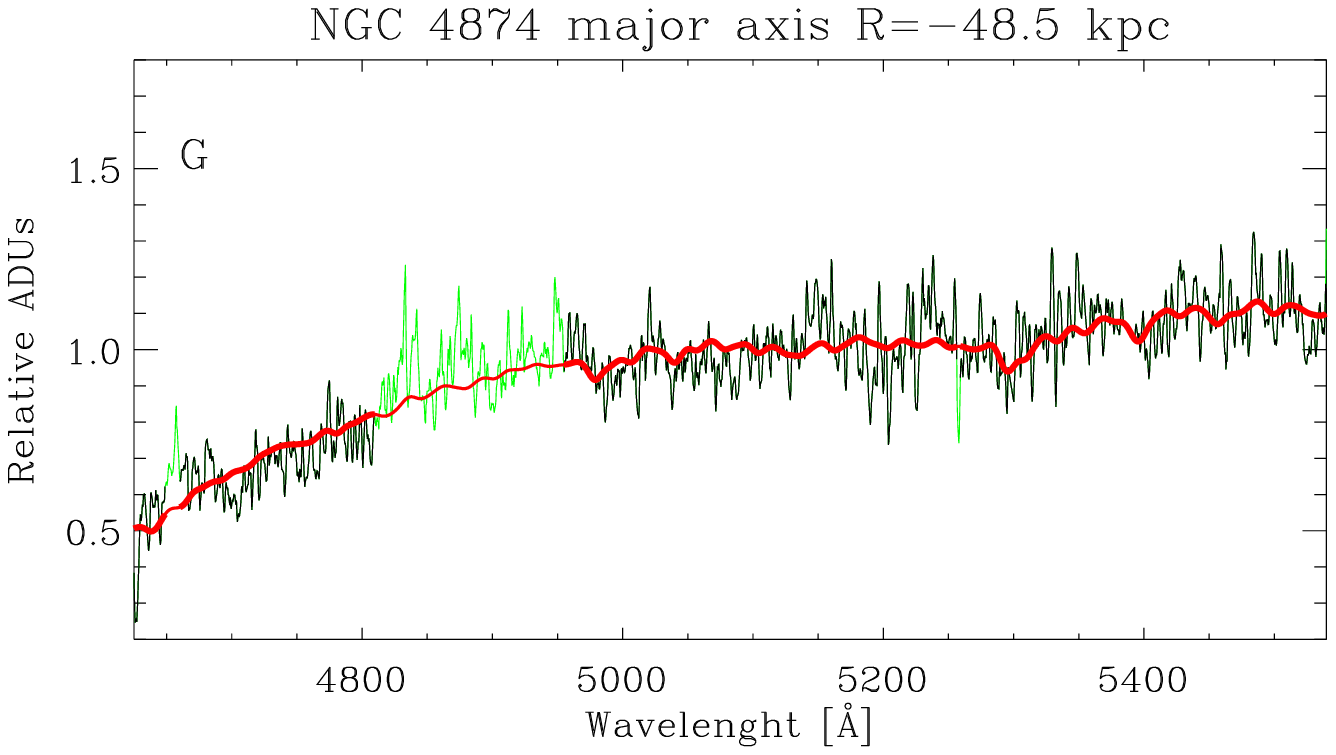,width=9.0cm,clip=}
\caption{{\bf -- Continued.}}
\label{fig:fit2}
\end{figure}

Summing adjacent spectra over part of the slit may affect the
absorption line shapes by either smoothing the intrinsic velocity
gradient or enlarging the absorption lines, if there is a substantial
velocity gradient in the sampled region.

Such an effect may be significant for run 1, where the slit
sampled halo light of galaxies whose systemic velocities differ by
$\sim 700$ \kms.
We estimated the maximum number of adjacent columns to be summed and
extracted in a one-dimensional spectrum to be approximately $\sim
30-40$, depending also on the CCD cosmetics. Within this range, the
maximum smoothing of the intrinsic velocity gradient is approximately
$\sim 20-30$ \kms, which is the same order as the measurement
errors. This value is obtained multiplying the number of columns by
the maximum velocity gradient along the slit \footnote{This gradient
  was computed from the velocity difference measured at the slit edges
  ($\sim 800$ \kms) divided by the number of pixels in the slit
  (960).}, which we estimated to be $\sim 0.8$ \kmsp.  For run 2, we
estimated the effect to be negligible, because no velocity gradient
was measured along the minor axis of NGC 4889 (Section
\ref{sec:stellar_kinematics}).

In the case of NGC 4874, the 1D spectra obtained from $40$
adjacent columns did not reach the required $S/N$. For those regions
we used the following iterative procedure:

\begin{enumerate}

\item{} We selected $N$ regions on the slit which are $M \leq30$ pixels
  wide. Each region corresponds to a two-dimensional spectrum, which
  we call ``stripe'' $S_i$ (with $i=1, ..., N$). Each of the $N$
  stripes is composed of $M$ one-dimensional adjacent
  spectra, with the wavelength direction oriented along columns.

\item{} We extracted the one-dimensional spectrum $S^C_i$, by summing
  the stripe along the slit direction, and assigned the luminosity
  weighted central position $R_i$ (in pixels) on the slit.  
  We associated thus a spectrum $S^C_i$ and a central
  position $R^C_i$ to each region of the slit.

\item{} We measured the radial velocity $V^C_i$ of the spectrum
  $S^C_i$. The radial velocity measurements are described in
  Section \ref{sec:stellar_kinematics}.

\item{} We selected a reference spectrum $S^C_{\rm ref}$. The other
  $N-1$ spectra were then shifted to the radial velocity $V^C_{\rm
    ref}$ of the reference spectrum. The $\Delta V$ applied to
  the $S^C_{i-th}$ spectrum was $\Delta V = \delta V \cdot (R^C_i -
  R^C_{\rm ref})$, where $\delta V$ (in \kmsp) is the (unknown) velocity gradient
  along the slit.
  The shifted spectra were then combined together to form the total
  spectrum $S^{\rm TOT}$ in the radial bin centered at $R^C_{\rm
    ref}$.
  $\delta V$ was determined via an iterative procedure that minimized
  the line widths in the combined spectrum $S^{\rm TOT}$.

\end{enumerate}

The above scheme and the iterative procedure for the determination of
$\delta V$ assumed a constant velocity gradient along the
slit.  In order to minimize systematics that may arise from this
assumption, we limited the radial binning to a maximum of 3 adjacent
stripes $S_{i-1,i,i+1}$, each of them 30 pixels wide at most.

\begin{figure}
\psfig{file=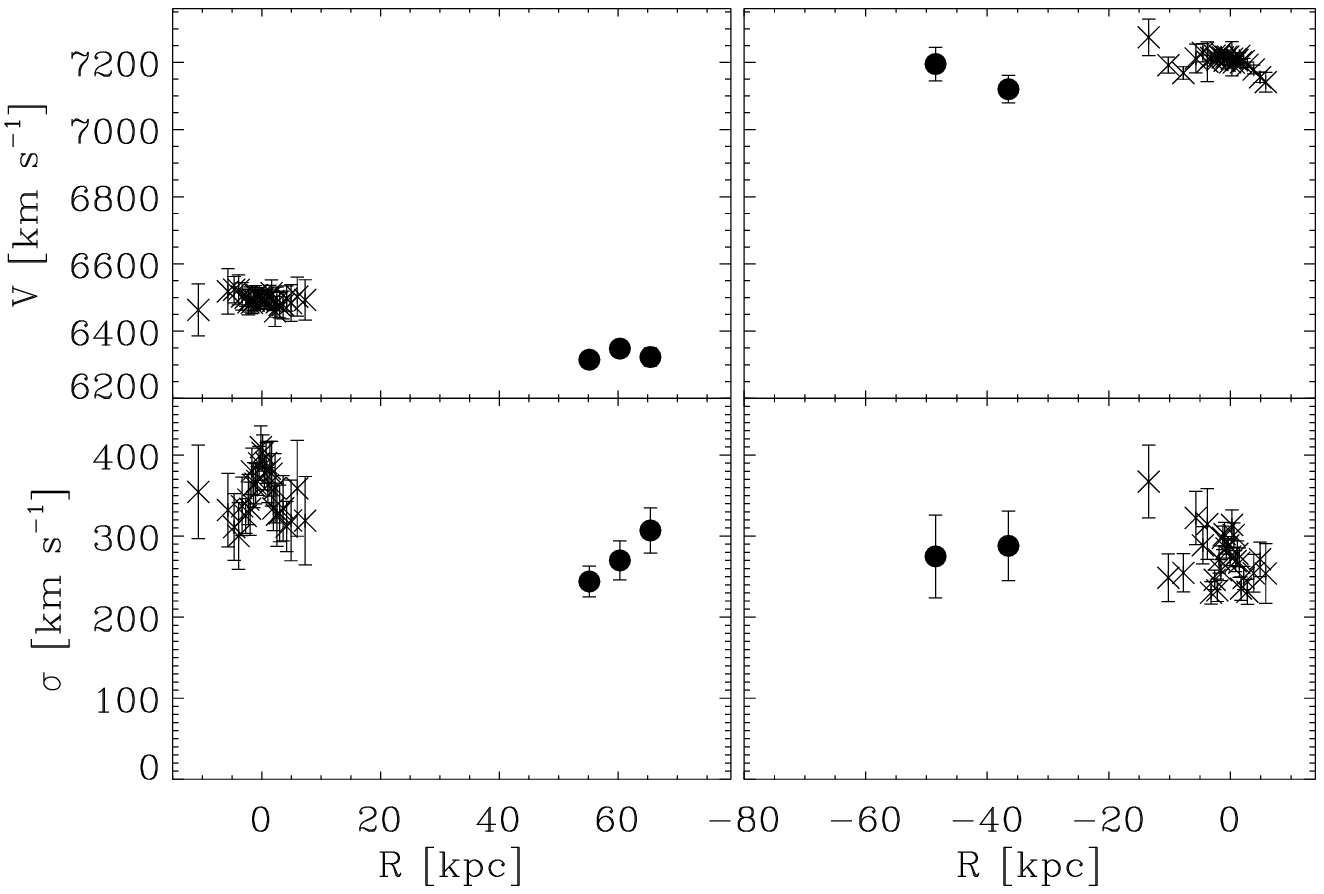,width=9.0cm,clip=}
\caption{Radial velocity ({\it upper panels}) and velocity dispersion
  profiles ({\it lower panels}) along the slit in run 1 (velocities
  are expressed in the heliocentric reference system). Measurements
  along the slit are divided in two panels, with {\it left panels}
  referring to the part of the slit close to NGC 4889 and {\it right
    panels} referring to the part of the slit close to NGC 4874. {\it
    Crosses:} data from \citet{Mehlert+00} shifted to the systemic
  velocities, 6495 \kms\ for NGC 4889 \citep{Moore+02} and 7205 \kms\
  for NGC 4874 \citep{Smith+04}. {\it Filled circles:} measurements
  from this paper for run 1. }
\label{fig:run1_kinem}
\end{figure}

\begin{figure}
\psfig{file=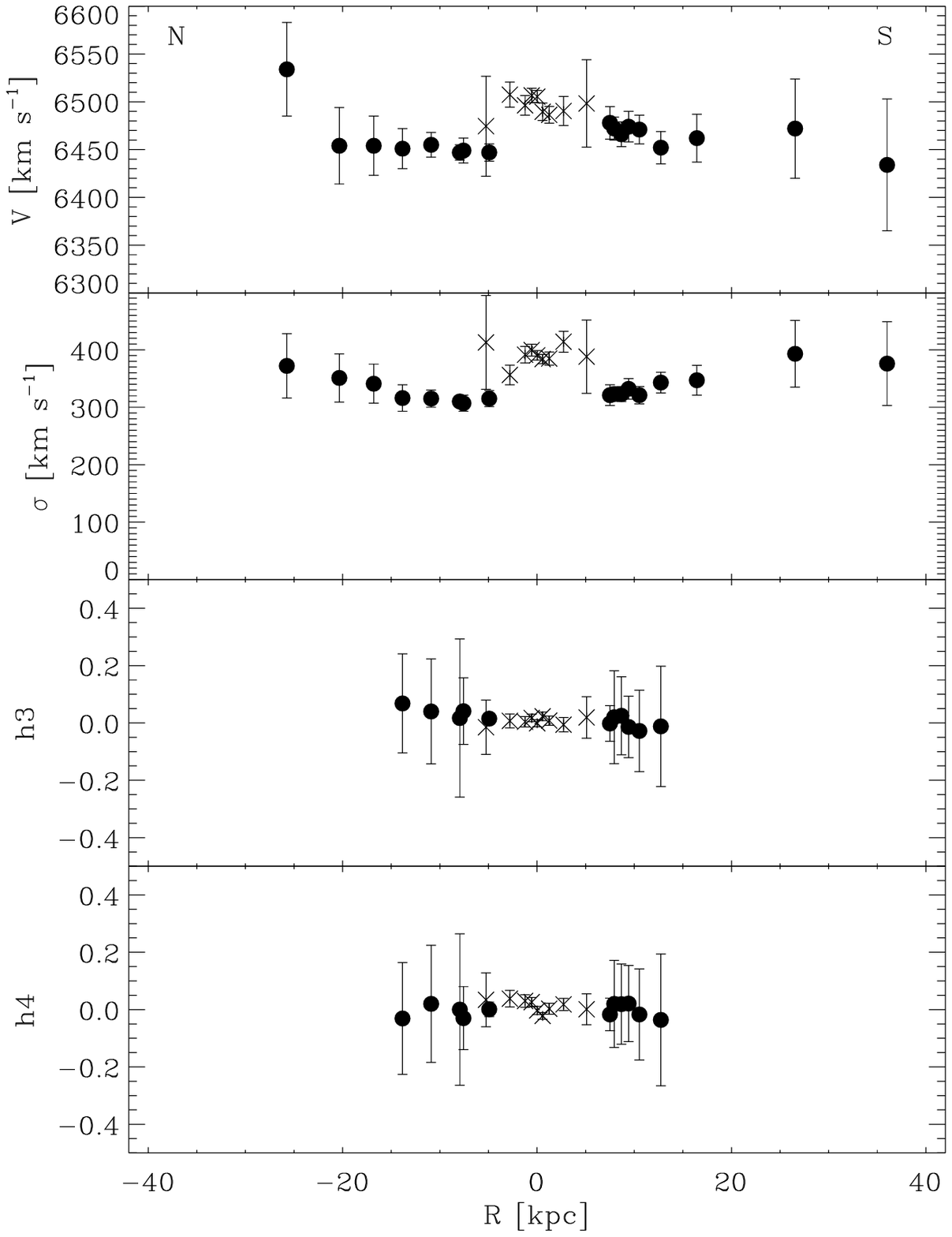,width=9.0cm,clip=}
\caption{Stellar kinematics along the minor axis of NGC 4889. {\it
    Crosses:} data from \citet{Corsini+08}, {\it filled circles:} 
  measurements from this paper for run 2.}
\label{fig:run2_kinem}
\end{figure}

\section{Stellar kinematics}
\label{sec:stellar_kinematics}

Stellar kinematics from the extracted spectra in the radial bins were
measured using the ``Penalized Pixel-Fitting method'' ({\small PPXF},
\citealt{Cappellari+04}). A library of kinematic template stars (MILES
library, \citealt{Sanchez-Blazquez+06}) was used together with
stellar templates observed during the observations.
The template spectra of the MILES library were convolved with a
Gaussian function to match the FOCAS spectral resolution.
We show some of the fit results in Figure \ref{fig:fit1}: the
  innermost and outermost spectra extracted along the minor axis of
  NGC 4889, the three outermost spectra along the major axis of NGC
  4889 and the two spectra of NGC 4874 are also shown.

The wavelength ranges used for the fitting are $4600 \AA <
  \lambda < 5550 \AA$ for run 1, and $4800 \AA < \lambda < 5850 \AA$
  for run 2.  Bad pixels coming from residual cosmic rays or emission
lines were properly masked and not included in the fitting procedure.

For each radial bin, the {\small PPXF} code built the optimal template
which  represents the observed spectrum by a linear combination of
different stellar templates. The use of a large spectral library and the
creation of one optimal template for each radial bin minimized errors
in the kinematics due to template mismatch.
In the outermost bins ($R \geq 60$ kpc in run 1 and $|R| > 20$ kpc in run
2) the construction of the optimal stellar template in {\small PPXF}
failed, because the absorption line features are less
  pronounced. The kinematics in these regions were measured using the
best template obtained from the nearest bin in which it could be
  otained. Referring to Fig. \ref{fig:fit1}, spectra in panels D, E, F
  and G were fitted using the optimal template determined when fitting
  the spectrum in panel C. The spectrum in panel B was fitted with the
  optimal template obtained fitting the spectrum at 16 kpc (not shown
  in the figure).
This approximation might introduce some systematic errors in the
kinematics of the outer bins due to template mismatch. Nevertheless,
in the radial bins ($|R|\leq 20$ kpc in run 2), we did not
observe significant differences in the kinematics when either the 
  ``central'' optimal template  at $R=-4.9$ kpc or the ``local''
optimal templates, found by {\small PPXF} for those bins, were
used. This  confirmed that errors on the kinematics caused by
template mismatch were negligible for the inner bins, and gave us
confidence that the template mismatch is not important in the
outer bins.  Errors on the measured kinematics were determined by
  means of Monte Carlo simulations, analyzing spectra generated from
  the best fit template model and adding the appropriate noise.

The measured radial velocities were then
shifted to the heliocentric reference system using the {\small
  MIDAS}\footnote{{\small MIDAS} is developed and maintained by the
  European Southern Observatory.} task {\small COMPUT/BARY}.

The fitted mean velocity, velocity dispersion and Gauss-Hermite
moments \citep{Gerhard93,vanderMarel+93}  and their errors are
shown in Figures \ref{fig:run1_kinem} and \ref{fig:run2_kinem} for
runs 1 and 2, respectively, and are listed in Table \ref{tab:results}.
 Gauss-Hermite moments were fitted only for spectra within
  $|R|<20$ kpc along the minor axis.

\section{Line strength indices}
\label{sec:indices}

Line strength indices were measured for NGC 4889 using observations
from both runs. The spectra of NGC 4874 did not have enough $S/N$ for
this.
We measured the \hb, \mguno, \mgdue, \mgb, \fei\ and \feii\
line-strength indices as defined by \citet{Faber+85} and
\citet{Worthey+94}. The spectra were convolved with a Gaussian
function to match the spectral resolution of the Lick system (FWHM =
8.4 \AA\ at 5100 \AA, \citealt{Worthey+97}). Three Lick standard stars
observed during the runs  were used to correct for offsets to the Lick
system. The measured offsets (FOCAS - Lick) are: 0.03 \AA\
  (\hb), $-0.30$ \AA\ (\feiii), $-$0.028 mag (\mguno), $-$0.016 mag (\mgdue), $-$0.037 \AA\
  (\mgb), $-$0.36 \AA\ (\fei), 0.024 \AA\ (\feii) and 0.024 \AA\ (\feiv).

\begin{figure*}
\hbox{
%/home/lcoccato/COMA/remove_emission_gandalf_like/sky_contribution/run1/mkplots.pro
 \psfig{file=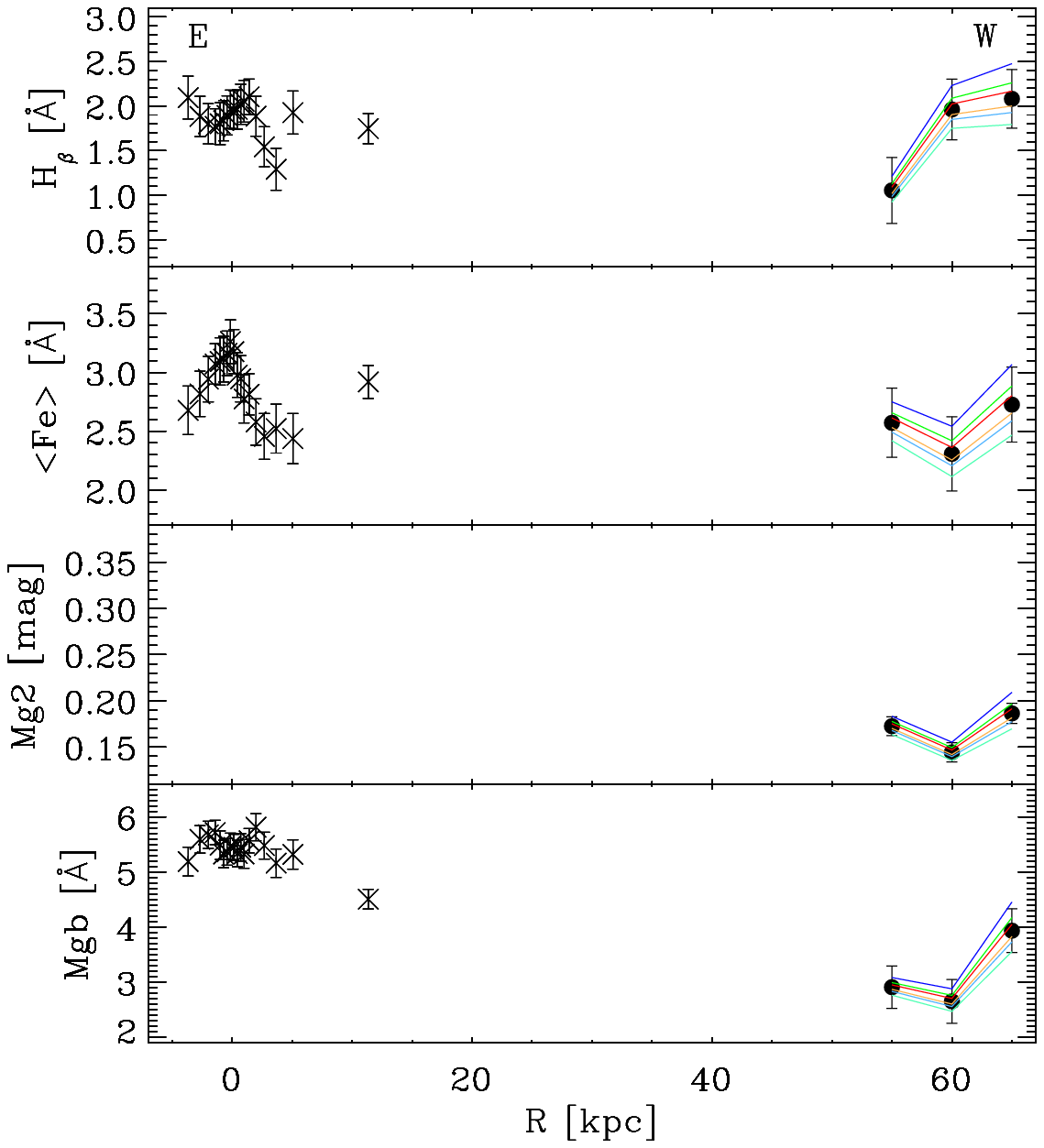,width=8.5cm,clip=}
%/home/lcoccato/COMA/remove_emission_gandalf_like/sky_contribution/run2/mkplots.pro
 \psfig{file=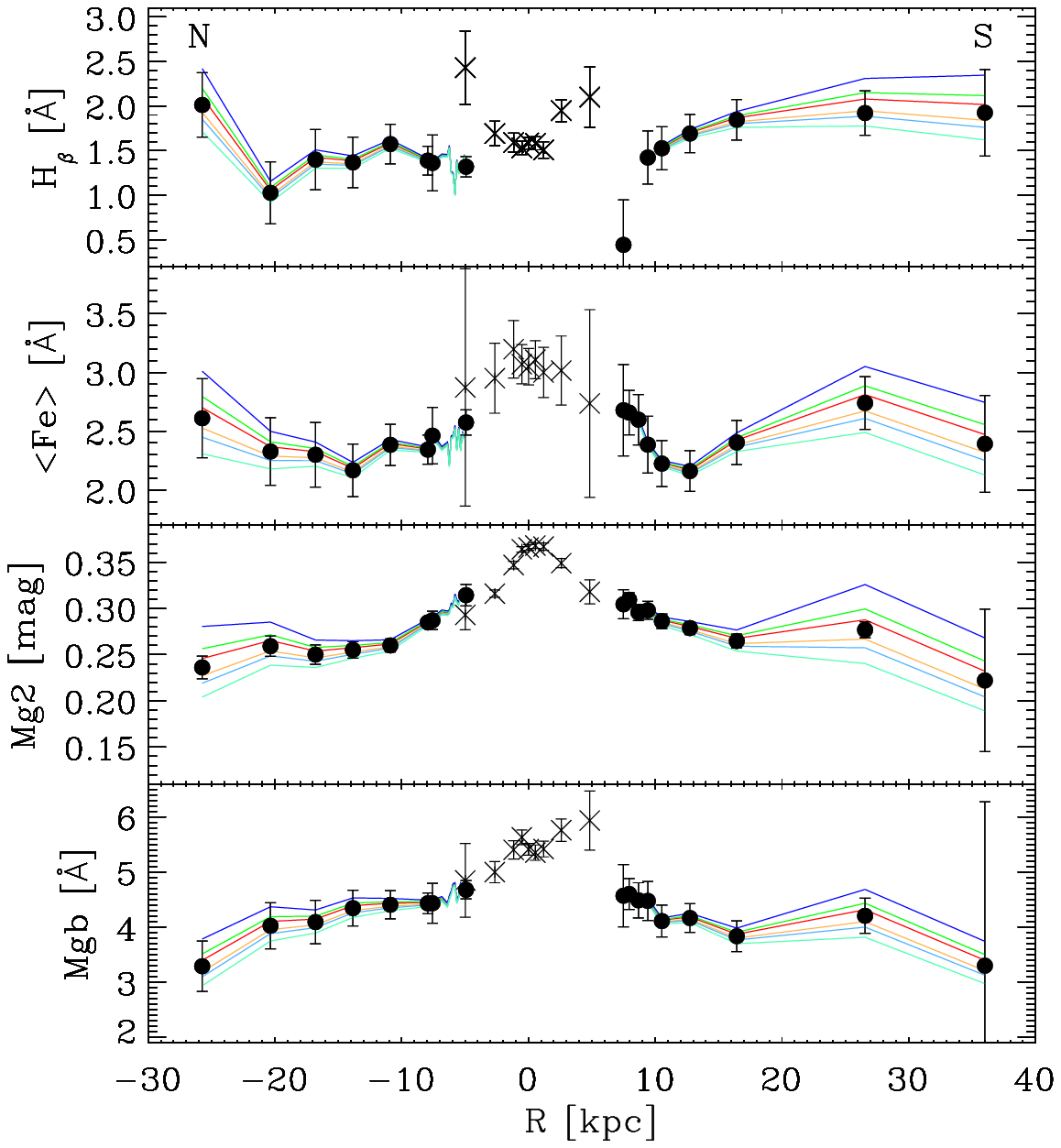,width=8.5cm,clip=}
}
\caption{Line-strength Lick indices for NGC 4889 measured along the
  major ({\it left panel}) and the minor axis ({\it
  right panel}).
Crosses: data from \citealt{Mehlert+00} (major axis) and
\citealt{Corsini+08} (minor axis).
{\it Filled circles:} measurements from this paper.
{\it Solid lines} indicate the systematic deviations of the
line-strength measurements caused by adding different residual sky
levels to the extracted galaxy spectrum. From top to bottom, the
residual sky levels we added are +2\% (i.e. sky subtraction is
underestimated by 2\%), +1\%, +0.5\%, $-0.5$\%, $-1$\%, $-2$\%
(i.e. sky subtraction overestimated by 2\%). See Section \ref{sec:sky}
for details. }
\label{fig:indices}
\end{figure*}

No focus correction was applied because atmospheric seeing
was the dominant effect during observations.

Errors on line-strength indices were determined via Monte Carlo
simulations, which accounted for the errors on radial velocity
measurement also. Errors from Monte Carlo simulations are
  consistent with those predicted by the analytic formulae of
  \citet{Cardiel+98}.

In the following we describe the details of additional corrections we
applied to the measurements in order to take into account 
instrumental effects. The final measured values are listed in
Table~\ref{tab:results} and  are shown in Figure~\ref{fig:indices}
together with the \citet{Mehlert+00} data along the major axis, and
the \citet{Corsini+08} data along the minor axis\footnote{We show the
  mean iron $<{\rm Fe}> = \frac{1}{2}\left( {\rm \fei} +{\rm \feii}
  \right)$ radial profiles and not the profiles of both iron
  indices to facilitate the comparison with the literature, {\bf where} 
  only $<{\rm Fe}>$ is provided.}.

\subsection{Correction for intrinsic broadening}
The line-strength measurements were corrected for the galaxy
intrinsic broadening following the procedure described in
\citet{Kuntschner+06}. Coefficients $C_I$ for each index
$I$ were determined as the ratio of i) the ``intrinsic'' values
($I_0$, in Angstrom) measured on the optimal stellar template; and ii)
the line-of-sight-velocity-distribution ``modified'' values
($I_{\rm LOSVD}$) measured on the best fit model (i.e. the optimal
template convolved with the galaxy LOSVD). Corrected galaxy line
strength indices (in \AA ngstrom) were then given by:

\begin{equation}
  I_{\rm corr} = C_I \cdot I_{\rm meas} = \frac{I_0}{I_{\rm LOSVD}} \cdot  I_{\rm meas} 
\end{equation}

\subsection{Correction for residual emission lines}
The long-slit spectra for the NGC 4889 minor axis showed strong
unresolved emission lines in the wavelength regions adjacent to the
Mg{\small I} bandpass and the \hb\ blue pseudo-continuum. 
These strong lines were at wavelengths that are not associated with
known galaxy emission lines or sky/auroral lines. Our conclusion was
 that they may be associated with internal reflections in the
spectrograph. To measure the spectral indices in the regions affected
by the unidentified emission lines, we simultaneously fit the galaxy
spectrum with a stellar template plus  Gaussian functions at the
wavelength of the detected emissions. The best-fit Gaussians were then
subtracted from the spectrum, to remove the observed emission lines.
As a further test, we checked the measured values of the line-strength
indices against the measurements on the best-fit stellar
template, convolved with the galaxy line of sight velocity
distribution determined in Section \ref{sec:stellar_kinematics}.
The two sets of values were consistent.

\subsection{Systematic effects caused by sky subtraction}
We then quantified the systematic errors in the line strength indices
measurements caused by sky subtraction . A residual sky contribution
of $\pm$0.5\%, $\pm$1\% and $\pm$2\% of the subtracted sky
spectrum\footnote{Residual sky contribution larger than 2\% would
  produce detectable sky features either in emission or absorption in
  the galaxy spectra. 2\% is also the maximum difference between the
  continuum of the sky measured at blank fields and the one measured
  simultaneously to galaxy observations (see Section \ref{sec:sky} and
  Fig. \ref{fig:sky_comparison} for discussion).} was added to the
extracted galaxy spectra, and then we performed measurements of the
line-strength indices on these modified spectra. Within the errorbars,
these new values were consistent with the measurements performed on the
best background subtracted galaxy spectra, but for the \mgdue\
index. This was probably due to the fact that the \mgdue\
pseudo-continua were defined on spectral regions not adjacent to the
index passband (unlike for the other spectral indices), therefore
variations of the sky level  have a larger effect on this index.

\section{Results and discussion}
\label{sec:discussion}

In NGC 4874 the measured stellar kinematics extend out to $\sim 50$
kpc ($\sim 1.4 R_e$) along its East side. Measurements of line
strength indices were not possible.

In NGC 4889 our measurements extend out to 65 kpc ($\sim 4.3 R_e$)
along its West side (close to its major axis) and $\sim 35$ kpc ($\sim
2.3 R_e$) along its photometric minor axis. This represents the
most spatially extended datasets with both stellar kinematics
{\it and} line strength indices for a brightest cluster galaxy.

\subsection{Kinematics}

{\bf NGC 4874.}  In NGC 4874 the outer mean velocity along the line of
sight is $<V> = 7150 \pm 37$ \kms, about $\sim 55$ \kms\ lower than
the galaxy systemic velocity ($V_{syst} = 7205 \pm 19$ \kms, as
reported by \citealt{Smith+04}). It is not straightforward to
associate it with rotation or tidal effects, given the errors on the
current measurements and the large uncertainties on the galaxy
systemic velocity. As examples, \citet[RC3 hereafter]{RC3} report
a systemic velocity of $7152 \pm 16$ \kms, while NED reports a
systemic velocity of $7224 \pm 11$ \kms.  The mean outer velocity
dispersion in NGC 4874 is $\sigma= 283 \pm 33$ \kms, which is
consistent with the central value found by \citet{Smith+04},
suggesting a flat velocity dispersion profile with radius.

\noindent {\bf  NGC 4889.}
In NGC 4889 the mean outer velocity is $<V> = 6326 \pm 44$ \kms, which
is $\sim 169$ \kms\ lower than the galaxy systemic velocity ($V_{syst}
= 6495 \pm 13$ \kms, as reported by \citealt{Moore+02}). The measured
velocity offset is significant when compared with measurements errors
and uncertainties on the systemic velocity. Although rotation in the
halos of ellipticals is not rare (e.g. \citealt{Rix+99, Coccato+08b}),
it is not straightforward to consider the velocity offset as evidence
of rotation, since we miss measurements on the East side of the
galaxy.
The outer mean velocity dispersion we measured in NGC 4889 is $\sigma=
266 \pm 17$ \kms, consistent with the outermost measurements ($\sim
10"$) by \citet{Moore+02}, revealing a flat velocity dispersion
profile outside the central drop.

\citet{Gerhard+07} developed a model for the core of the Coma cluster
whereby the two BCGs, NGC 4874 and NGC 4889, are in the process of
merging, tidally stripping each others' halos to create an elongated
distribution of intracluster light in which they are embedded. It is
worth noting that, out to the radii probed by the present
observations,
the velocity dispersion profiles of NGC 4874 and NGC 4889 are nearly
constant and indicative of stars still bound to the central
galaxy. This is consistent with recent velocity dispersion profiles
measurements based on planetary nebulae and globular clusters for
local brightest cluster galaxies (e.g. \citealt{Coccato+08b};
\citealt{Doherty+09}; \citealt{Proctor+09}; \citealt{Schuberth+09};
McNeil et al. 2010 A\&A submitted). These galaxies do not show an
increase in the velocity dispersion profile as observed in other BCGs
(e.g NGC 6166 in \citealt{Kelson+02}), which was interpreted as
generated by stars not bound to the galaxy but free floating in the
cluster potential.

\begin{figure}
\psfig{file=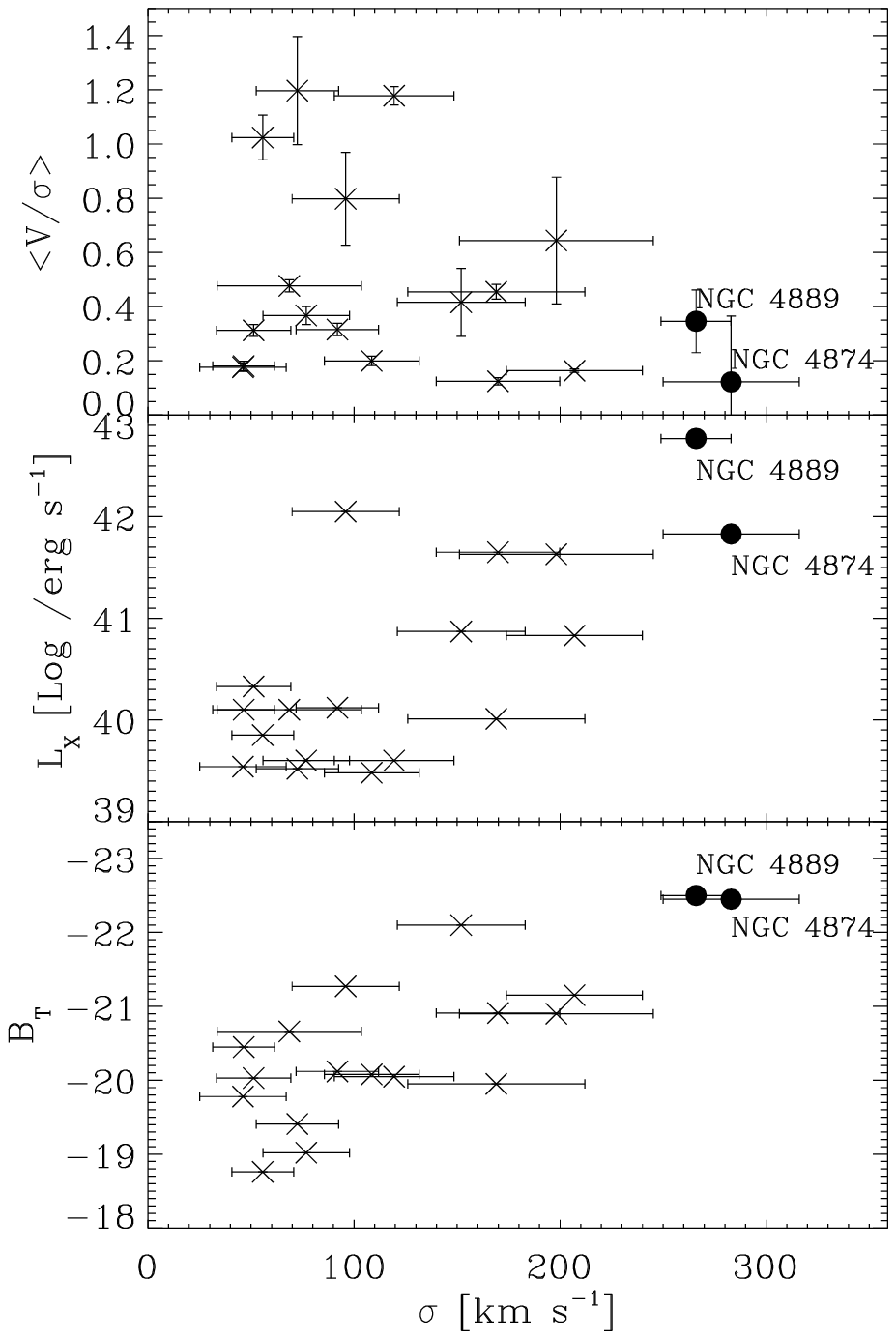,width=8.5cm,clip=}
\caption{Correlations between the outermost measured value of the halo
  velocity dispersion and mean $<V/\sigma>$ of the halo ({\it upper
    panel}), total X-ray luminosity ({\it central panel}) and total
  luminosity in the B band ({\it lower panel}). {\it Crosses}
  are data from \citet{Coccato+08b}, {\it filled circles} are data
  from this paper.}
\label{fig:correlations}
\end{figure}

We can thus compare the kinematic properties of the outer regions of
NGC 4874 and NGC 4889 with the kinematic properties of the outer halos
of other early-type galaxies.  To do this, we refer to the results of
\citet{Coccato+08b} based on radial velocities of Planetary Nebulae in
the halos of early-type galaxies. Their sample covers the range 50
\kms\ $\lesssim \sigma \lesssim$ 220 \kms\ of velocity dispersion in
the outer halo. The two galaxies studied in this paper fall in the
upper $\sigma$ range of the galaxy sample explored by
\citealt{Coccato+08b} and this enables us to extend their $\sigma$
range. Following \citet{Coccato+08b}, in Figure \ref{fig:correlations}
we compare the halo velocity dispersion with other galaxy properties,
$<V/\sigma>$ (we assume that the observed velocity offset is due to
rotation), total X-ray luminosity ($L_X=6.8\cdot10^{41}$ erg s$^{-1}$
for NGC 4874 and $L_X=5.9\cdot10^{42}$ erg s$^{-1}$ for NGC 4889, from
\citealt{Pellegrini05}) and total $B$-band magnitude ($B_T=-22.45$ for
NGC 4874 and $B_T=-22.50$ for NGC 4889, from RC3).  Both NGC 4874 and
NGC 4889 follow the same trend found in other early-type galaxies,
therefore supporting the earlier statement that the stars in their
halos are still bound to the inner galaxies in the radial range
explored in this work.

\subsection{Line strength indices}

In NGC 4889, the measured set of Lick indices provides information in
the radial interval from $\sim 7$ to $\sim 65$ kpc, and is
complementary to data available for the central regions
\citep{Mehlert+00,Corsini+08}. 

In the region of overlap the current data set and the published values
agree along the minor axis ($|R|\sim5$ kpc), but for \hb.  

In the central parts of the galaxy, the mean velue from
  \citet{Corsini+08} data is higher than the inward extrapolation of
  our data by 0.3 \AA. Their outermost central value differs from ours
  of 1.1 \AA\ at the same radius ($R\sim -5$ kpc, see
  Fig. \ref{fig:indices}).

The difference could be related either to the limited number of
  stars we used to correct the FOCAS observations to the Lick system,
  or to sky subtraction, affecting Corsini's measurements (see Section
  \ref{sec:indices}). But in these cases we would expect a discrepancy
  also in other indices, and not only in \hb.  Another reason could
be the presence of emission lines, but neither in our spectra nor in
Corsini's spectra \hb\ emission lines were detected at 5 kpc.
Therefore, the reason for the discrepancy remains unclear.

Values of \hb\ range from 0.5 to 2.3 \AA. Line strength indices
related to \mgi\ and $<{\rm Fe}>$ show declining profiles, from the
central values of \mgdue\ $\sim 0.37$ \AA, \mgb\ $\sim 5.5$ \AA\ and
$<{\rm Fe}>$ $\sim 3.4$ \AA, to outer values of \mgdue\ $\sim 0.15$
\AA, \mgb\ $\sim 3$ \AA\ and $<{\rm Fe}>$ $\sim 2.5$ \AA.

In \citet{Coccato+10} the combined set of line strength
indices for NGC 4889 is used to derive age, metallicity and
alpha-enhancement for single stellar population models and their
radial gradients from the center out to 60 kpc. There we find
evidence for a change in the stellar population gradient and
content, suggesting a different evolutionary history between the
galaxy and its halo.
We refer to this forthcoming paper for a detailed discussion of the
implications of these measurements for the properties of the stellar
populations in NGC 4889, and for the constraints they put on the
galaxy formation mechanism.

\begin{acknowledgements}
      
  The authors wish to thank T. Hattori for the professional support
  during the observations and for providing the FOCASRED package used
  in the data reduction; R. Bender, E. Iodice, H. Kuntschner,
  L. Morelli, R.P. Saglia, D. Thomas and J. Thomas for useful
  discussions.

  This research has made use of the NASA/IPAC Extragalactic Database
  (NED) which is operated by the Jet Propulsion Laboratory, California
  Institute of Technology, under contract with the National
  Aeronautics and Space Administration, and of the ``Penalized
  Pixel-Fitting method'' by Cappellari \& Emsellem (2004).

\end{acknowledgements}

\bibliography{14476_cl}

\begin{landscape}

\begin{table}
\centering
\caption{Stellar kinematics and line strenght indices for NGC 4874 (run 1) and NGC 4889 (runs 1 and 2).}
\begin{tabular}{l c  c  c  c  c  c  c  c c  c  c c}
\hline \noalign{\smallskip}
  $R$    &  $V_{\odot}$  &$\sigma$   &$h_3$                 & $h_4$               &    \hb                      &\feiii  & \mguno              & \mgdue                            & \mgb                               & \fei &  \feii  &  \feiv \\
  (kpc) & (\kms)        &(\kms)      &                     &                     &    (\AA)                    & (\AA)  &   (mag)             & (mag)                             & (\AA)                              & (\AA)& (\AA)   &  (\AA)\\
  (1)   &   (2)         &  (3)       &  (4)                &   (5)               &    (6)                      &   (7)  &   (8)               &  (9)                              & (10)                               & (11) & (12)    &  (13)\\
\hline 
\noalign{\smallskip}
{\bf Run1: }\\
\noalign{\smallskip} 
{\bf N 4874}\\ 
 $-48.50$&7195 $\pm$ 50 &275 $\pm$ 51&$  -   $             &$   -  $             &--                           &  $-$           &   $-$               &   $-$                             &   $-$                              &   $-$           &   $-$           &  $-$    \\
 $-36.51$&7120 $\pm$ 41 &288 $\pm$ 43&$  -   $             &$   -  $             &--                           &  $-$           &   $-$               &   $-$                             &   $-$                              &   $-$           &   $-$           &  $-$     \\
 {\bf N 4889}\\
 55.17  & 6315 $\pm$ 18 &244 $\pm$ 19&$  -   $             &$   -  $             &1.05        $\pm$ 0.37       &  $-$           & 0.05  $\pm$  0.08  & 0.173   $\pm$     0.010           &  2.91  $\pm$    0.39               &  2.43  $\pm$    0.38 &  2.71  $\pm$    0.46& $-$  \\
 60.31  & 6348 $\pm$ 24 &270 $\pm$ 24&$  -   $             &$   -  $             &1.96        $\pm$ 0.34       &  $-$           & 0.02  $\pm$  0.09  & 0.144    $\pm$     0.010          &  2.65  $\pm$    0.40               &  2.12   $\pm$   0.43 &  2.50  $\pm$    0.46&  $-$ \\
 65.46  & 6323 $\pm$ 27 &307 $\pm$ 28&$  -   $             &$   -  $             &2.08$^{(*)}$ $\pm$ 0.33      & $-$             & 0.03  $\pm$  0.09  & 0.186    $\pm$      0.011         &  3.93   $\pm$   0.40               &  2.81   $\pm$   0.41 &  2.64  $\pm$    0.50&  $-$ \\
\noalign{\smallskip}
{\bf Run2:}\\
\noalign{\smallskip} 
 {\bf N 4889}\\   
$-25.74$& 6534 $\pm$ 49 &372 $\pm$ 56&$  -   $             &$   -  $             &   2.01 $\pm$     0.36       &    --          & --                  &0.24$^{(*)}$  $\pm$ 0.01           & 3.29$^{(*)}$ $\pm$    0.46          & 2.44  $\pm$    0.46 &  2.78  $\pm$    0.49  & 1.40 $\pm$   0.46 \\ 
$-20.34$& 6454 $\pm$ 40 &351 $\pm$ 42&$  -   $             &$   -  $             &   1.03 $\pm$     0.35       &4.27 $\pm$ 0.87 & --                  &0.26$^{(*)}$ $\pm$  0.01           & 4.02$^{(*)}$  $\pm$   0.42          & 2.38   $\pm$   0.40 &  2.27   $\pm$   0.42  & 1.32 $\pm$   0.37 \\     
$-16.80$& 6454 $\pm$ 31 &341 $\pm$ 34&$  -   $             &$   -  $             &   1.40 $\pm$     0.34       &3.80 $\pm$ 0.72 & --                  &0.25$^{(*)}$ $\pm$  0.01           & 4.09$^{(*)}$  $\pm$   0.39          & 2.22   $\pm$   0.38 &  2.38  $\pm$    0.39  &   --              \\  
$-13.84$& 6451 $\pm$ 21 &316 $\pm$ 23&$  0.07$ $\pm$ 0.17&$ -0.03$ $\pm$ 0.20&   1.37 $\pm$     0.28       &3.76 $\pm$ 0.70 & 0.16  $\pm$  0.07   & 0.255  $\pm$  0.009               &  4.34  $\pm$    0.33               &  2.12  $\pm$    0.31 &  2.21 $\pm$     0.32 & 1.28 $\pm$   0.33 \\                   
$-10.89$& 6455 $\pm$ 13 &315 $\pm$ 15&$  0.04$ $\pm$ 0.18&$  0.02$ $\pm$ 0.20&   1.57 $\pm$     0.22       &4.25 $\pm$ 0.58 & 0.17  $\pm$  0.05   & 0.260  $\pm$  0.007               &  4.41  $\pm$    0.26               &  2.44  $\pm$    0.24 &  2.33  $\pm$    0.25 & 1.25 $\pm$   0.27 \\  
$ -7.94$& 6447 $\pm$  8 &310 $\pm$  9&$  0.02$ $\pm$ 0.28&$  0.00$ $\pm$ 0.26&   1.39 $\pm$     0.16       &4.50 $\pm$ 0.46 & 0.18  $\pm$  0.04   & 0.285  $\pm$  0.005               &  4.43   $\pm$   0.19               &  2.34  $\pm$    0.17 &  2.34  $\pm$    0.18 & 1.39 $\pm$   0.21 \\  
$ -7.57$& 6449 $\pm$ 13 &307 $\pm$ 14&$  0.04$ $\pm$ 0.12&$ -0.03$ $\pm$ 0.11&   1.36 $\pm$     0.31       &4.53 $\pm$ 0.33 & 0.18  $\pm$  0.08   & 0.287  $\pm$  0.010               &  4.43   $\pm$   0.36               &  2.42   $\pm$   0.33 &  2.50 $\pm$     0.35 & 1.44 $\pm$   0.15 \\                           
 $-4.92$& 6447 $\pm$  9 &315 $\pm$ 14&   0.02  $\pm$ 0.02&   0.00  $\pm$ 0.03&   1.32 $\pm$     0.12       &4.89 $\pm$ 0.16 & 0.19  $\pm$  0.11   & 0.314  $\pm$  0.012               &  4.68   $\pm$   0.17               &  2.62  $\pm$    0.20 &  2.53  $\pm$    0.13 & 1.57 $\pm$   0.07 \\  
    7.50& 6478 $\pm$ 17 &321 $\pm$ 18&$ -0.00$ $\pm$ 0.06&$ -0.02$ $\pm$ 0.06&   0.44 $\pm$     0.50       &4.59 $\pm$ 0.51 & 0.19  $\pm$  0.12   & 0.305  $\pm$  0.016               &  4.57   $\pm$   0.57               &  2.88  $\pm$    0.53 &  2.47 $\pm$     0.58 & 1.73 $\pm$   0.24 \\                           
    7.94& 6472 $\pm$ 12 &323 $\pm$ 12&$  0.02$ $\pm$ 0.16&$  0.02$ $\pm$ 0.15&   --                        &     --         & 0.19  $\pm$  0.06   & 0.309  $\pm$  0.008               &  4.60   $\pm$   0.28               &  2.66  $\pm$    0.26 &  2.65 $\pm$     0.28 & 1.73 $\pm$   0.45 \\                           
    8.68& 6466 $\pm$ 13 &323 $\pm$ 13&$  0.03$ $\pm$ 0.14&$  0.02$ $\pm$ 0.14&   --                        &4.67 $\pm$ 0.49 & 0.18  $\pm$  0.07   & 0.296 $\pm$   0.009               &  4.49   $\pm$   0.32               &  2.55  $\pm$    0.29 &  2.64 $\pm$     0.31 & 1.77 $\pm$   0.22 \\                           
    9.42& 6474 $\pm$ 16 &332 $\pm$ 18&$ -0.01$ $\pm$ 0.11&$  0.02$ $\pm$ 0.13&   1.42 $\pm$     0.30       &4.90 $\pm$ 0.56 & --                  & 0.30$^{(*)}$ $\pm$ 0.01          &4.48$^{(*)}$  $\pm$  0.35            &  2.42   $\pm$   0.33 &  2.36 $\pm$     0.36 & 1.83 $\pm$   0.25 \\   
   10.53& 6471 $\pm$ 15 &321 $\pm$ 15&$ -0.03$ $\pm$ 0.14&$ -0.02$ $\pm$ 0.16&   1.53 $\pm$     0.24       &4.48 $\pm$ 0.63 & --                  & 0.29$^{(*)}$ $\pm$ 0.01          &4.11$^{(*)}$  $\pm$  0.29            &  2.26  $\pm$    0.27 &  2.19  $\pm$    0.29 & 1.50 $\pm$   0.28 \\   
   12.74& 6452 $\pm$ 17 &343 $\pm$ 18&$ -0.01$ $\pm$ 0.21&$ -0.04$ $\pm$ 0.23&   1.69 $\pm$     0.21       &4.76 $\pm$ 0.51 & --                  & 0.28$^{(*)}$ $\pm$ 0.01          &4.16$^{(*)}$  $\pm$  0.26            &  1.95   $\pm$   0.24 &  2.37 $\pm$     0.25 & 1.61 $\pm$   0.23 \\   
   16.44& 6462 $\pm$ 25 &347 $\pm$ 26&$ -    $             &$ -     $            &   1.85 $\pm$     0.23       &4.61 $\pm$ 0.44 & --                  & 0.27$^{(*)}$ $\pm$ 0.01          &3.83$^{(*)}$   $\pm$ 0.29            &  2.08   $\pm$   0.26 &  2.73 $\pm$     0.28 & 1.44 $\pm$   0.21 \\   
   26.55& 6472 $\pm$ 52 &393 $\pm$ 58&$ -    $             &$ -     $            &   1.92 $\pm$     0.25       &      --        & --                  & 0.28$^{(*)}$ $\pm$  0.01          &4.20$^{(*)}$  $\pm$  0.32            &  2.03   $\pm$   0.31 &  3.44 $\pm$     0.33 & 1.56 $\pm$   0.23 \\   
   36.01& 6434 $\pm$ 69 &376 $\pm$ 73&$ -    $             &$ -     $            &   1.92 $\pm$     0.49       &     --         & --                  & 0.22$^{(*)}$ $\pm$  0.08          &3.30$^{(*)}$  $\pm$  2.98            &  2.02   $\pm$   0.57 &  2.77 $\pm$     0.60 &      --           \\   
\noalign{\smallskip}                   
\hline                                 
\end{tabular}                          
\label{tab:results}                 
\begin{minipage}{24cm}                 
  Notes -- Measurements marked by $^{(*)}$ were obtained by removing
  an emission line component from the spectrum.  The symbol -- means
  that the measurement has not been performed.
\end{minipage}                         
\end{table}

\end{landscape}

\end{document}